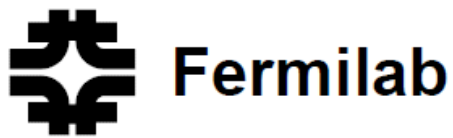

# Shielding studies for superconducting RF cavities at Fermilab[1*]

Camille Ginsburg and Igor Rakhno

Fermilab, Illinois, 60510, Batavia, USA

---

[1]Work supported by Fermi Research Alliance, LLC under contract DE-AC02-07CH11359 with the U.S. Department of Energy.

[*]Presented at 10th Workshop on Shielding Aspects of Accelerators, Targets and Irradiation Facilities (SATIF-10), CERN, Geneva, Switzerland, 2-4 June, 2010.

# Shielding studies for superconducting RF cavities at Fermilab

**Camille Ginsburg and Igor Rakhno**

Fermilab, Illinois, 60510, Batavia, USA

### Abstract

*A semi-empirical method that allows us to predict intensity of generated field emission in superconducting RF cavities is described. Spatial, angular and energy distributions of the generated radiation are calculated with the FISHPACT code. The Monte Carlo code MARS15 is used for modeling the radiation transport in matter. A comparison with dose rate measurements performed in the Fermilab Vertical Test Facility for ILC-type cavities with accelerating gradients up to 35 MV/m is presented as well.*

### Introduction

Test facilities for high-gradient superconducting RF cavities usually are strong sources of γ-radiation due to field emitted electrons inside the cavities. The field-induced emission is generally the result of various imperfections, *e.g.,* residual dust contamination, in the cavity. The imperfections can give rise to a significant enhancement of local electric field and, consequently, field emitted electrons which generate gammas in the surrounding material. The neutron component generated by the gammas is not negligible for accelerating gradients of above ~20MV/m. The design of shielding for such facilities involves significant uncertainties because of the lack of a reliable model of the field emission.

We use a semi-empirical approach to predict the radiation source term in the superconducting RF cavities. It is based on realistic spatial, angular and energy distributions of field emitted electrons modeled with the FISHPACT code [1] as well as dose rate measurements performed at the DESY Tesla Test Facility (TTF). Modeling of the interaction with and transport in matter for the generated radiation is performed with the Monte Carlo code MARS15 [2]. A comparison between predicted and measured dose rates for the Fermilab Vertical Test Facility is performed.

### Source Term Issues

#### *Field-induced dark current in superconducting RF cavities*

Experimental observations of field-induced emission (in other words−dark current) in superconducting RF cavities can be summarized as follows: (i) surface imperfections can happen anywhere, but field emission occurs mostly around irises−locations with the highest local electric field; (ii) for a given superconducting RF cavity the emission usually does not occur at several sites; it usually happens at a single site and lasts until a significant amount of RF energy stored in the cavity is lost to the generated dark current. Therefore, in our model we focus on regions around the irises, and the field emitted electrons are modeled inside the cavity until they hit the cavity surface. Phase-space coordinates of such events represent a digitized source term for subsequent modeling of secondary particle generation and transport in the entire system with the MARS15 code.

Dark current generation is assumed to be equiprobable for all regions around the irises. For a given region, however, the probability of field-induced emission depends greatly on the magnitude and RF phase of the surface electric field [3]. Therefore, the relative probabilities of possible electron trajectories differ significantly, and the most probable trajectories usually do not correspond to the highest electron energy gain in the accelerating field. In our model, the field emission is generated with azimuthal symmetry.

### *Trajectory analysis*

A realistic two-dimensional model for the radiation source term was developed to describe the trajectories and energy distributions of field emitted electrons generated in superconducting RF cavities at high accelerating gradients. The FISHPACT code used to model the electron trajectories is interfaced with the POISSON SUPERFISH code [4], a simulation package used to calculate RF electromagnetic fields. Although the simulation provides a field emission current, given input field emission parameters, only the electron trajectories and energies have been used here. The dose estimated in the simulation using standard parameters from literature [3], was found to be substantially higher than justified by existing data, so data have been used to normalize the predicted dose, as discussed later.

The cavity cell structure and simulated surface electric field are shown as a function of cavity Z in Figs. 1 and 2. Electrons emitted from iris regions may be accelerated along the cavity axis and acquire significant energy. An example of simulated trajectories for an emission site, in which electrons can reach energy almost as high as the cavity accelerating gradient, is shown in Fig. 3.

**Figure 1: The surface electric field of a 9-cell cavity (solid, normalized to 1 MV/m gradient), and the cavity cell structure (dashed) as a function of cavity Z from a SUPERFISH simulation. The electric field peaks in the cavity iris regions.**

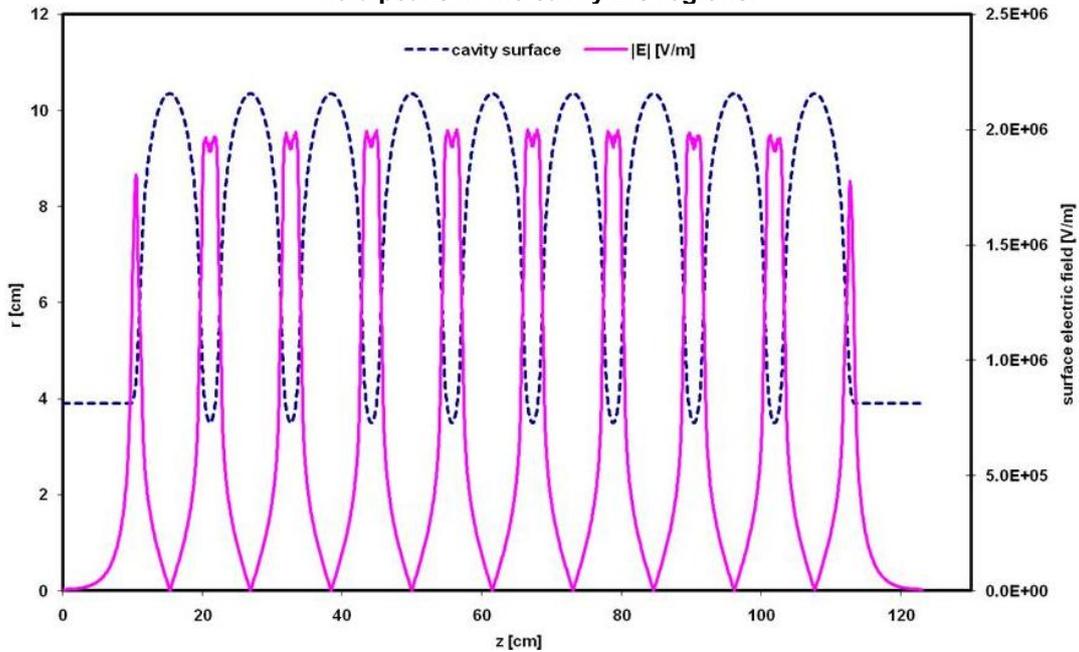

**Figure 2: A zoomed view of the previous figure, showing detail around an iris in a 9-cell RF cavity.**

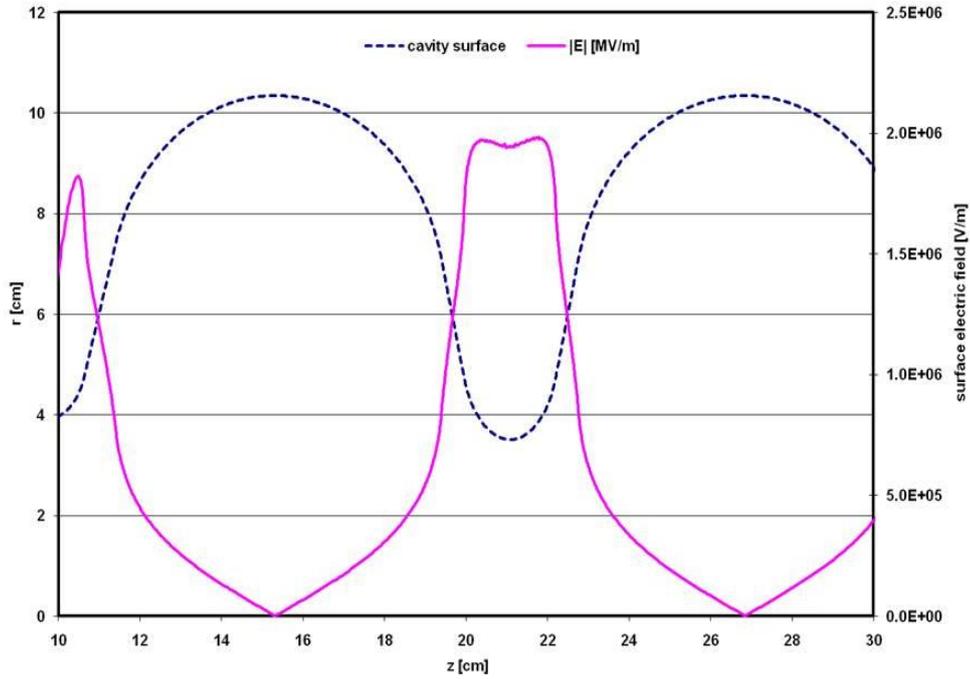

**Figure 3: Simulated electron trajectories generated in a 9-cell SRF cavity with an accelerating gradient of 30 MV/m. The curves inside the cavity correspond to electron trajectories for 10 degree increments in the RF phase, for the half period in which the electric field has the correct sign to pull electrons from the surface.**

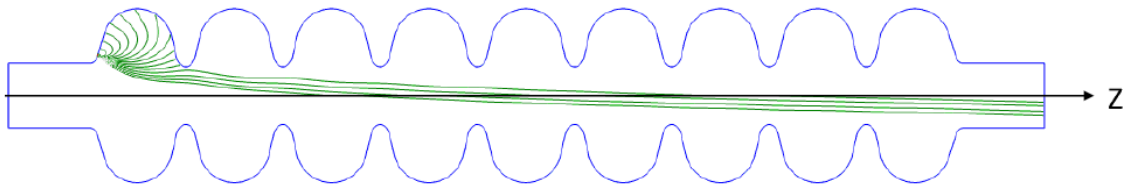

The maximum gradient of 30 MV/m has been chosen to correspond to the largest gradient at which field emission is likely to result in a substantial dose rate immediately above cryostat, as determined from data. In addition, this is approximately the largest gradient for which the Fermilab Vertical Test Facility must be able to test typical cavities without interrupting the RF system with radiation monitoring system trips.

For every single iris, five emission sites were studied: the iris itself and four sites in its vicinity (two on each side). Emission sites near cavity beampipes were taken into account as well. The surface electric field has the right sign to facilitate the electron quantum tunneling under the surface potential barrier for 50% of the test time. The corresponding RF phase intervals were divided into 18 10° bins, so that 17 sample trajectories with electric field greater than zero are generated for each emission site. The phase-space coordinates ($\mathbf{r}$, $\mathbf{E}$, $\Omega$) of electrons when their trajectories hit the inner cavity surface were recorded for a subsequent modeling with the MARS15 code.

### *Fowler-Nordheim model*

Fowler and Nordheim [3] were the first to provide a description of field emission as quantum mechanical tunneling:

$$j \approx A(E)E^2 \exp\left(-\frac{B(E)\phi_e^{3/2}}{E}\right), \quad (1)$$

where $j$ is the density of the generated electron dark current, $E$ is the local electric field, $A$ and $B$ are slowly varying functions of $E$, and $\varphi_e$ is work function of the emitting material. In practice, however, the field induced emission was observed at much smaller fields than those compatible with the Equation (1). In order to better fit experimental results, the following expression was found to be more accurate:

$$j \approx SA(E)\beta^2 E^2 \exp\left(-\frac{B(E)\phi_e^{3/2}}{\beta E}\right). \quad (2)$$

Two new parameters are introduced in Equation (2): (i) the effective emitting area factor, $S$, and (ii) the local field enhancement factor, $\beta$. Equation (2) implies that the emission occurs at small localized spots around imperfections and the local field at the imperfections can be significantly higher (by a factor of about 100 and more) than that predicted for an idealized cavity surface. The two parameters vary substantially and are usually determined from experimental data.

### *Intensity of the predicted dark current*

Because of the variability in field emission parameters, experimental data from DESY/TTF are used to normalize the predicted dark current, as shown in Fig.4. From these data, we determined that for 90% of all measurements the dose rate measured under the external shielding did not exceed 5 R/hr.

**Figure 4: Measured maximum dose rates at DESY/TTF under the external shielding.**

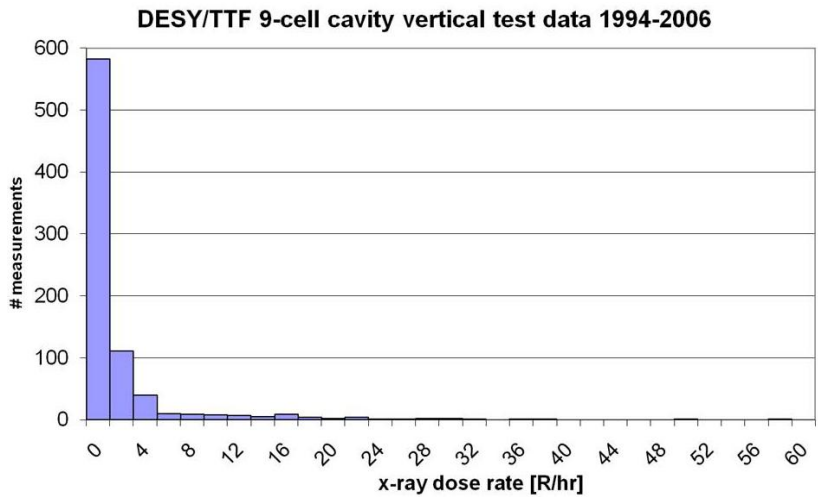

Therefore, we assume that the total dark current generated in a superconducting RF cavity corresponds to the dose rate of 5 R/hr predicted at a similar location in Fermilab Vertical Test Facility, after controlling for differences in radiation shielding. Practically speaking, if the field emission is too high, cavities are removed from the test facility and undergo an extra cleaning procedure.

**Vertical Test Facility at Fermilab**

The Vertical Test Facility for superconducting RF cavities at Fermilab has been in operation since 2007. The facility currently consists of a single vertical test cryostat VTS1. Radiation shielding for VTS1 was designed for single 9-cell 1.3 GHz cavities using a very simple field emission model as the radiation source [5]. The shielding consists of three parts: a plug internal to the cryostat, a movable shielding lid above the cryostat, and the concrete walls of the pit containing the cryostat. The internal shielding consists of a cylindrical assembly above the cavity containing layers of lead, steel and borated polyethylene. The external shielding consists of a concrete/steel movable shielding lid and borated polyethylene in instrumentation feedthrough regions. The RF system is interlocked through a radiation monitor outside of the shielding to maintain a controlled area designation.

Two additional cryostats with common design, VTS2&3, are being procured, and are sized such that six 9-cell cavities can be installed per cryostat. The test throughput will be gained through common cool-down and warm-up time, with cavities tested sequentially. Space for additional shielding, either internal or external to the cryostat, is limited. An evaluation of the radiation shielding was performed, to minimally extend the VTS1 shielding design to a six-cavity configuration in VTS2&3. The external shielding for VTS2&3 was unchanged. The internal shielding consists of cylindrical lead blocks above each of the cavities and cylindrical layers of steel and borated polyethylene above the upper cavities. The configuration of internal shielding with respect to the cavities is shown in Fig.5. Other cryostat components also serve as radiation shielding: (i) layers of copper and G10 above the internal shielding and under the top plate; (ii) the steel top plate; (iii) several cylindrical shells around the superconducting RF cavities – magnetic shield of Cryoperm-10 with aluminum support liner, helium vessel, copper thermal shield and steel vacuum vessel. Various small components such as cables and pipes are not included in the model. The superconducting RF cavities are tested in a superfluid helium bath at 2 K. Measurements at VTS1 revealed that total cool-down time is about 180 minutes, and that cooling the internal shielding occupies about a third of that time. The estimated total cool-down time for VTS2&3 without the six internal shielding lead blocks, initially inherited from the VTS1 design, is about 240 minutes. Keeping all the six lead blocks would increase the cool-down time by about 30%. Using an extra layer of external shielding on the movable lid, if necessary, is preferable in this case, and the extra mechanical load is expected to be well within the shield's load tolerance.

For VTS1, measured dose rates are shown in Fig. 6. According to the described model, the predicted dose rate under the external shielding is < 0.250 R/hr 90% of the time. The measurements are well within this value, to within the limited statistics. The described approach to shielding design for test facilities is justified by the possibility of extra cleaning procedures for tested superconducting RF cavities.

**Figure 5: A fragment of a model VTS2 or VTS3 insert with six RF cavities, showing a lead block above each cavity and a steel/borated polyethylene block above the entire cavity assembly.**

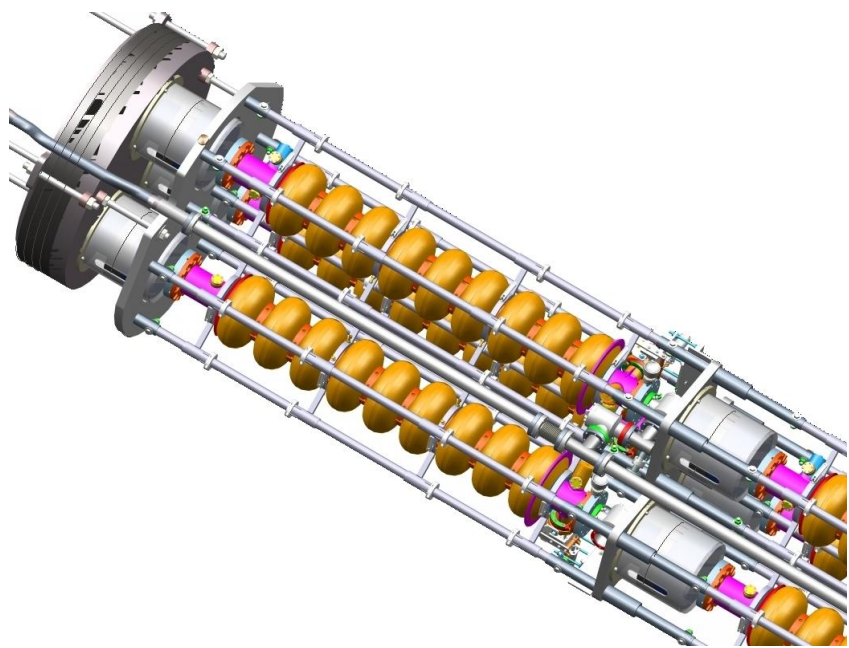

**Figure 6: Dose rates measured at VTS1 for 9-cell superconducting RF cavities with various accelerating gradients.**

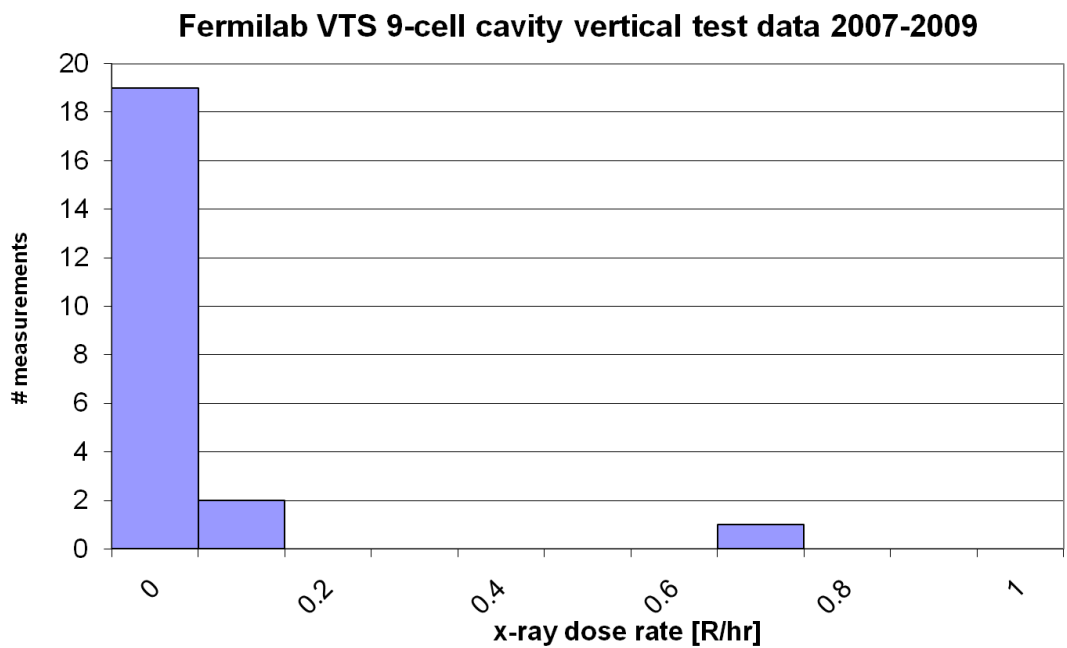

**Acknowledgements**

Assistance from G. Wu, C. Reid, J. Ozelis and C. Sylvester (FNAL), and W.-D. Möller and P.-D. Gall (DESY) is gratefully acknowledged.